# ARTIFICIAL INTELLIGENCE: FRAMEWORK OF DRIVING TRIGGERS TO PAST, PRESENT AND FUTURE APPLICATIONS AND INFLUENCERS OF INDUSTRY SECTOR ADOPTION

Richard Fulton[1], Diane Fulton[2] and Susan Kaplan[3]

[1]Department of Computer Science, Troy University, Troy, Alabama, USA
[2]Department of Management,
Clayton State University, Morrow, Georgia, USA
[3]Modal Technology, Minneapolis, Minnesota, USA

## ABSTRACT

*To gain a sense of the development of Artificial Intelligence (AI), this research analyzes what has been done in the past, presently in the last decade and what is predicted for the next several decades. The paper will highlight the biggest changes in AI and give examples of how these technologies are applied in several key industry sectors along with influencers that can affect adoption speed. Lastly, the research examines the driving triggers such as cost, speed, accuracy, diversity/inclusion and interdisciplinary research/collaboration that propel AI into an essential transformative technology.*

## INTRODUCTION

Artificial Intelligence (AI) is an evolving science and art. Developments come in flashes and spurts over time. The scientific community changes its focus on different topics and applications. Technological developments can and will continue to expand the problem solving and innovative capabilities of AI. Researchers build on what has been done in the past, implement in the present and dream about what can happen in the future. Together, these developments over time lead to the state of the art of a technology like AI.

This paper presents a time-evolving Framework for AI (FAI) based on past and present adoptions and future expectations of technology uses. Triggers such as cost, speed, accuracy, customization, inclusivity/ diversity and cross discipline/collaboration are factors that push an organization to adopt and transform a new technology. When there are dramatic changes in the environment, what the customer needs, competitiveness in the industry and increased resources to implement a new technology, these become influencers in how rapidly technology becomes transformational as well. In this framework, the state of the art of AI is impacted by triggers, influencers and time. Three distinct industrial sectors including agriculture, education and healthcare illustrate the sector-dependent nature of AI application development over time, spanning the past, present and future. The authors conclude with an in-depth discussion of the six driving triggers of AI transformative technology adoption.

## RELATED WORKS

This section briefly reports the most related work to examining the triggers and influencers of AI technology adoption over time based upon a variety of theories and research models. The first group of theories and models pertinent to the development of artificial intelligence include those related to technology acceptance and adoption. The 3 most used acceptance/adoption models are the Technology Acceptance Model (TAM), Diffusion of Innovations Theory (DOI) and the Unified Theory of Acceptance and Use of Technology (UTAUT) [1].

TAM, the most widely tested empirical model, proposes three technology acceptance factors including 1) "perceived usefulness", 2) "perceived ease of use" and 3) "attitude towards use" and focuses on the individual [2]. In contrast, DOI focuses on both individuals and organizations and four factors of time, channels of communication, social systems and innovation which impact technology diffusion and adoption [3]. Differences in adopter characteristics in the DOI model categorize firms and individuals within firms as early adopters, innovators, laggards, late majority and early majority leading credibility to industry sector differences discussed in this paper [3].

Lastly the UTAUT model is a compilation model built on 8 models (including TAM and DOI) emphasizing effort expectancy, performance expectancy, social influence, and facilitating conditions [4]. "Facilitating conditions" mean removal of barriers impeding technology adoption. For example, using the UTAUT model in applying new technology in e-learning, "facilitating conditions" included providing financial resources, new infrastructure, additional human resources and innovative educational content [5]. These "facilitating conditions' are included in this study under the "influencer" construct subcategory *resources*.

Both TAM and DOI models use the constructs of "perceived usefulness" and "relative advantage" [6]. A new construct of "perceived benefits of technology adoption" incorporated into the International Technology Adoption (ITA) model combined the previous constructs of technology utility to the individual with benefits to the company's well-being and corresponds closely to the "triggers" of *speed*, *accuracy*, *cost*, and *customization* and the "influencers" of *competitive advantage* and *customer needs* presented in this paper.

Digital transformation in industry is a compelling topic and focus of a framework called "The Digital Transformation Journey" [7]. In their framework, the compelling construct of "mounting challenges and drivers" means finding ways to use technology to do business in new and better ways [7]. Coronavirus (Covid-19) in late 2019, for example, is considered a pressure point or "driver" of technology transformation in a variety of industrial sectors [7].

An example in the healthcare sector of a "driver" or "influencer" of AI technology is the coronavirus in Wuhan, China in 2019 which used AI tools to provide early detection of the coronavirus, isolating those areas with the virus [8]. It is likely the experience of a global pandemic will have a long-lasting and global impact on AI diffusion finding new ways of early detection which will help prevent future pandemics and influence health policies worldwide [8].
Increased competition is another "challenge" creating market pressure that if not addressed, can lead to loss of market share and revenues [7]. The Framework of AI uses the construct of "influencers" of *changing environments* and *competitive advantage* which correspond to transformational "challenges and drivers".

In addition to building on previous model constructs of "benefits" such as this research's "triggers" which add value, usefulness and benefits to the individual (less time to do a task) and organization (reducing costs and mistakes) and "influencers" of *changing environments*, *resources* and *competitive advantage*, the authors enhance the existing theories and models by adding two new "triggers of AI technology adoption" – *diversity/inclusion* and *crossdiscipline/collaboration* in their framework. Not addressing these essential issues could sabotage transformational adoption of AI.

In fact, the more leaders understand the biases in technology [8] and the need for collaboration across disciplines/fields, the better they can improve its usefulness and therefore, increase its transformational adoption [9]. Lastly, this research supports the idea that AI technology is a dynamic phenomenon which changes over time and a better understanding of technology changes through past, present and future developments can help increase individual and organizational ability to build their AI maturity [10].

## ARTIFICIAL INTELLIGENCE FRAMEWORK

Successes in one industry spur interest in another sector. Some sectors are quick to adopt new technological applications such as AI and others are more cautious. Factors that can prompt or influence adoption include changing environments such as climate change or a pandemic event like the 2019 coronavirus (covid-19) or evolving customer needs for a product or service [11][12][13].

Often, organizations are searching for competitive advantage such as cost, quality, or better satisfying a particular niche of consumers. For example, a recent McKinsey study showed advanced AI adopter firms were 52% more likely to increase their market share by 52% and 27% had growth in their marketplace compared to those who were testing or moderately implementing AI [14]. Lastly, there are changing priorities in the allocation and budgeting of resources depending on societal expectations and organizational readiness [15]. Figure 1 gives a schematic of the Framework for AI (FAI), from development triggers to adoption influencers based on past, present and future AI technology trends.

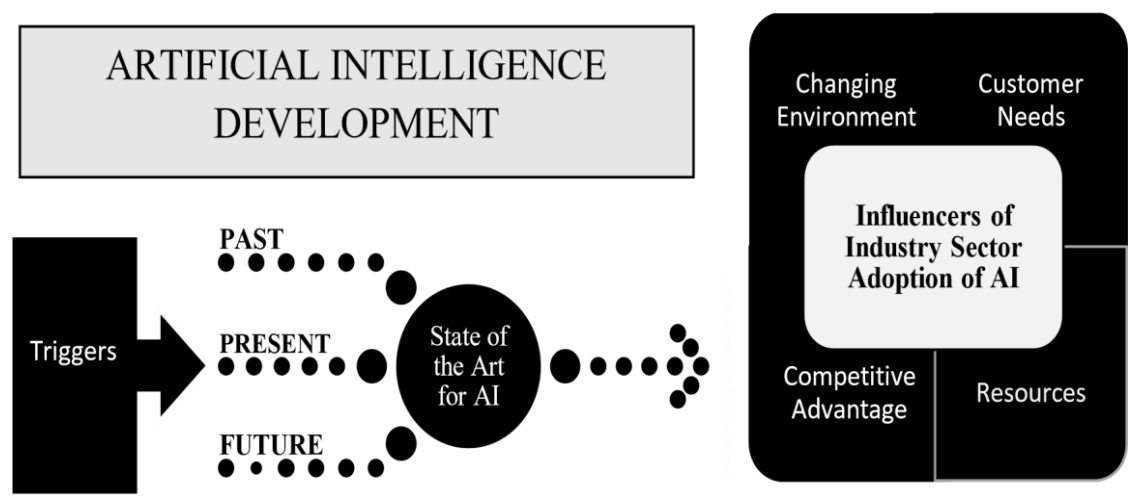

Figure 1. Framework for Artificial Intelligence: Triggers to Past, Present and Future Application Development and Influencers of Industry Sector Adoption

**Past**

AI began in the 1940s, demonstrating that a new form of computing was possible, with an approach derived from known cognitive processes and neurobiology. The initial purpose of AI was to automate, through computers, non-analytical human knowledge, from symbolic computation processes, connectionist ones, or a combination of both. AI was initially considered a branch of computer science with limited application and restricted by the capabilities of the hardware of the time.

Turing, a British mathematician, developed a code breaking computer called the *Bombe* in the early 1940's that successfully broke the Enigma code used by the Germans during World War II, a task thought impossible by most human mathematicians at the time. He also developed the Turing Test, that states "if a human is interacting with another human and a machine and unable to distinguish the machine from the human, then the machine is said to be intelligent" [16].

In 1956, John McCarthy offered one of the first and most influential definitions of AI: "The study is to proceed on the basis of the conjecture that every aspect of learning or any other feature of intelligence can in principle be so precisely described that a machine can be made to simulate it" [17].

One of the most famous AI examples is IBM's Deep Blue chess playing program, which beat the world chess champion Gary Kasparov in 1997. This expert system processed 200 million possible moves per second and determined the optimal next move looking 20 moves ahead [18].

**Present**

The current definition of artificial intelligence (AI) has transformed into "computing systems that are able to engage in human-like processes such as learning, adapting, synthesizing, selfcorrection and use of data for complex processing tasks" [19].

AI has become a vital element for the development of many services and industrial sectors in the 21st century. This discipline of computer science studies algorithms to develop computer solutions that copy the cognitive, physiological, or evolutionary phenomena of nature and human beings. The data, examples of solutions, or relationships between these facilitate the resolution of diverse problems [20]. AI exhibits, in certain aspects, "an intelligent behavior" that can be confused with that of a human expert in the development of certain tasks [21].

The Deep Blue project inspired the development of Watson, a computer that was able to beat the two best Jeopardy Game players in the world in 2011. Its software could process and reason using natural language, and draw from a massive supply of information poured into it in the months before the competition [22].

At present, AI has been redirected towards the construction of solutions to problems analyzing large volumes of data which change over time. Currently, the systems for approaching functions using iterative techniques, and the neural network architectures interconnected with each other, make up most of the techniques, which are grouped under the terms "Machine Learning" and "Deep Learning".

AI is becoming a growing presence in our society. From the intelligent sensors that make a car drive autonomously to mobile assistants, we are already surrounded by AI in some way or the other at all times [23]. Alexa, Siri, Cortana, security surveillance, fitness/dieting apps and online customer service are all examples of AI [24]. A large portion of the global population use these products/services in their everyday lives and the demand and popularity are ever growing [24].

**Future**

AI is a game changing technology and disruptor. Within 10 years, it is predicted 375 million workers will need to change occupations as a result of widespread use of AI [24]. AI and machine learning are predicted to reshape most sectors but particularly manufacturing, energy, transportation, agriculture, labor markets, and financial management [25].

AI will not only impact our personal lives but also fundamentally transform how organizations make decisions and interact with employees and customers. One of the most vital questions will be how AI systems and humans can coexist with each other. Which decisions should be made by AI, which ones by humans, and which ones in collaboration will be an issues all companies need to address in the future [22].

## KEY SECTOR APPLICATIONS: AGRICULTURE

**Past**

Agriculture is a sector that includes studies in science, engineering, and economics. The deductive techniques of AI expert systems have been used in the field of agriculture to integrate crop management which encompassed irrigation, nutritional problems and fertilization, weed control-cultivation, herbicide application, and insect control/insecticide application. Additional subject areas were plant pathology, salinity management, crop breeding, animal pathology, and animal herd management [26].

Agricultural applications of expert systems and decision support systems have also benefited the simulation of processes and the management of supply operations [27][28].

In other studies, AI has been used in quality control processes, whether or not they are supported by artificial vision [29] or in processes of justification of food policy decisions, such as when the use of AI is analyzed as a collaborative tool between the different actors that supply the agri-food chain, using distributed computing processes [30].

In the field of science, climate aspects are studied through modeling and solar radiation is predicted using neural networks [31][32].

**Present**

Interest in the application of AI to the world of agriculture and its multiple facets has been growing in recent years as it has proven to be a powerful tool for data analysis [33].

Current AI technology investigates the price behavior of agri-food products [34][35][36]. In these cases, artificial neural networks and machine learning techniques are applied to investigate the price variations of agricultural commodities.

The expansion and intensification of industrial and technological agriculture have increased production, decreased the number of people suffering from poor nutrition and ensured richer and more resource-intense diets around the world. Industrial agricultural activities also generate employment, improve economic growth and boost the service sector in industrial regions [37].

Agriculture 3.0 brought robotics and automation to the agricultural world, as evidenced by agricultural machinery that performs complete cycles of agricultural work such as planting, spraying, and harvesting [38][39][40][41].

**Future**

Now, agriculture 4.0, combines intelligent farms and the interconnection of machines and systems, and seeks to adapt production ecosystems by optimizing the use of resources such as water, fertilizers, and phytosanitary products. In addition, it uses big data and imaging technology to arrive at "precision agriculture" [42][43][44][45].

Combined with genetic engineering and the use of data, it can solve an important part of agriculture by maximizing efficiency in the use of resources and adapting to climate change and other challenges [46]. To this end, the use of big data in decision-making is essential [47][48]. The technification of agriculture, decision support systems and the inclusion of concepts of Industry 4.0 by agri-food companies will continue to generate increased innovation in AI [49].

## APPLICATIONS IN EDUCATION SECTOR

**Past**

The IBM supercomputer Watson was watched across school and university campuses and all were delighted with the computer besting the 1994 world chess champion. In 2011, Watson with its victory in the game show *Jeopardy* against the two highest winners, heralded the era of cognitive computing with its potent natural language processing, knowledge representation and reasoning capabilities.

The educational interest in AI was initially captured through computers playing games but early versions of educational tutorials, learning management systems, simulations and iterative computer learning in the 1900s and early 2000s started the AI revolution in education [50][51][52].

**Present**

Universities have been particularly impacted by the 2019 coronavirus pandemic due to the inperson nature of traditional education. They are responding to this threat by investing in digital technologies such as cloud, AI, analytics, immersive learning spaces, and digital curricula. In fact, more than 80% of institutions are allocating over 25% of their 2021 IT budgets toward digital initiatives [53].

"Customization of learning has been happening through rising numbers of adaptive education programs, gaming, and software. These systems are personalized by enabling repeated lessons that students haven't mastered, and generally helping students to work at their own pace, space and liberty" [23].

Individualized automated tutoring has been developed to help students to learn easily and on their own schedules [54]. At Colorado State University, online students and tutors are using AI powered by Cognii, an Edtech company, to improve learning and assessment tools [55].

Another recent example of AI advancement is AlphaGo—a software or 'machine learning' developed by DeepMind, the AI branch of Google—that was able to defeat the world's best player at Go, a very complex board game considered more difficult than chess [56]. The AlphaGo program proved that the computer and deep learning can reach new heights and further advance human understanding in certain topics.

'Machine learning' is a subfield of artificial intelligence that includes software able to recognize patterns, make predictions, and apply the newly discovered patterns to situations that were not included or covered by their initial design.

**Future**

AI has the potential to modify the quality, quantity, delivery, and nature of education. It also promises to change forever the role of parents, students, teachers, and educational systems. Using Artificial Intelligence systems, software and support, students can learn from across the world at any time. These kinds of applications are taking the place of certain types of classroom instruction and may replace teachers in some cases [23].

AI can contribute to changing education via the automation of administrative teaching tasks, software programs that favor personalized education, the detection of topics that need reinforcement in class, the guidance and support of students outside the classroom, and the use of data in an intelligent way to teach and support the students [57].

Three techniques of AI are particularly relevant for future educational developments – personalization systems (knowledge and individualized adaptation of the student), software agents (intelligent programs and robots with autonomy and the ability to learn) [58] and ontologies and semantic web [59].

When developed and applied in education, these systems and techniques can be powerful resources for improving the teaching–learning process, since they are able to generate a kind of virtual teacher who is fully trained and has human characteristics, yet is able to interact ubiquitously (that is, at any time and place) [54].

By harnessing the power of AI and deep learning, educators can gain insights from the vast quantities of data collected from their students, make better decisions and improve student retention. Teachers can access detailed feedback on how learners are processing information. Big data can help answer key online learning questions—what are the most ideal ways to teach complex ideas and which parts of a course are best taught in person instead of online. Big data helps students find the right courses; customize them to their needs and keep them on the right track [55].

Most EdTech products will have an AI or deep learning component in the future. AI could help online learners self-assess, increase connectivity in global classrooms and create social simulation. Limitations include the uncertainty of how humans learn and fears among faculty that they must be retrained or could be displaced completely [55].

"Remote learning will coexist with on-campus education. As institutions accelerate their focus on student diversity and address unique educational needs, it is critical for them to make necessary technological investments to support their teaching models" [53].

In the future, higher educational institutions should expand outreach by using online courses and digitization of content to enable on-demand access by students across different geographies for remote learning, self-directed learning or specialized skill development. Secondly, increase funding to facilitate online learning, particularly enhancing IT capabilities – cloud platforms, collaborative tools, data security measures, AI bots and assessments. Lastly, educational organizations must learn to mine data assets and use AI's analytical solutions to develop personalized content, upskill faculty and enable remote proctoring, communications and virtual assistants [53].

## APPLICATIONS IN MEDICAL/HEALTHCARE SECTOR

**Past**

A recent review of the history of clinical decision support states the dramatic improvement in the medical sector due to the advent of cognitive aids to support diagnosis, treatment, care-coordination, surveillance and prevention, and health maintenance or wellness [60][61].

Some studies highlighted the importance of AI in healthcare, especially in medical informatics but there is still work to be done on examining the impacts and consequences of the technology [61][62].

**Present**

In the medical profession, image recognition tools are already outperforming physicians in the detection of skin cancer [63]. Molecular imaging modalities have also been effective in diagnosing neurodegenerative diseases [64].

Digital medicine and wearable devices are presently used in healthcare by mining data for anomaly detection, prediction, and diagnosis/decision making. Wearable devices and sensors have been used to continuously track physiologic parameters which guided patient care strategy that improved outcomes and lowered healthcare costs in cardiac patients with heart failure [65]. They also have been effective to improve diagnosis and management in neurological disorders such as Parkinson's disease [66].

Machine learning applications in healthcare have been helpful in earlier disease detection and prediction. For example, machine learning models were used in identifying stable subsets of predictive features for autism behavioral detection and blood biomarkers for autism [67][68].

Machine-learning algorithms were also used in the prediction of periventricular leukomalacia in neonates after cardiac surgery [69].

**Future**

Deep learning for automated and/or augmented biomedical image interpretation will continue to be used in radiology, pathology, dermatology, ophthalmology and cardiology with strict protocols and benchmarks in place to ensure data integrity and fairness. However, sensor-based, quantitative, objective and easy-to-use systems for assessing many diseases has the potential replace traditional qualitative and subjective ratings by human interpretation in the future [70].

Future AI in healthcare must be able to use machine learning to handle structured data such as images, data, genetic data, and natural language processing to mine unstructured texts. Then it must be trained through healthcare data before it can assist physicians with disease diagnosis and treatment options [71].

AI in medicine will continue with informatics approaches from deep learning information management to control of health management systems, including electronic health records, and active guidance of physicians in their treatment decisions. Also in the future, healthcare will increase its use of robots to assist elderly patients and targeted nanorobots, a unique new drug delivery system [72].

## TRIGGERS

Certain factors are accelerating the growth and use of AI throughout our society and will continue to be triggers for AI's transformative impact. AI can be used as a competitive strategy in all economic sectors particularly in cost/pricing advantages, customizing or personalizing products and services, and research using data mined from present and potential customers.

In addition, many AI advances have been accomplished by finding ways to increase the speed and accuracy of data resources and data research which can accelerate innovations while increasing the level of quality for consumers. Lastly, in our pursuit of the positive contributions of AI, we must be mindful of creating products and services that appeal to an inclusive and diverse group of people. Another way to increase the potential of AI is to use collaboration and reach across disciplines and sectors. Please see Figure 2 for the critical triggers impacting AI.

### Speed

Artificial intelligence systems can take control of many factors in an organization. For example, in an educational classroom – AI can control time-consuming tasks like accounting processes, record keeping, filling out forms, producing documents and automatically grade assignments freeing up time for teachers to improve the quality of learning, increase active learning and help students when needed [73].

In a survey about the benefits of AI in the workforce, 61% of respondents said it helped them have a more efficient and productive workday [74]. Almost half (49%) felt it improved their decision-making and accelerated time to insights, while 51% said they believed AI enabled them to achieve a better work/life balance [75].

The three highest rated tasks to benefit from AI adoption were: 1) understanding trends and patterns; 2) moving data from one place to another and 3) accessing data residing in different places across the organization [74].

### Accuracy

It is also predicted that 70% or more of companies will use some type of Artificial Intelligence in their operations because AI builds efficiency and effectiveness [24].

In the healthcare field, for example, AI can use sophisticated algorithms to 'learn' features from healthcare data, which can bring about insights for clinical practice and because it can be equipped with learning and self-correcting abilities, will improve its accuracy based on feedback over time [71].

In a recent interview with Susan Kaplan, the VP of a high-tech firm called Modal Technology located in Minneapolis, she cited that the "joint venture partnership between Modal Technology and medical researchers and scientists at McGill University Health Center Research in Montreal, Canada, using a new and mathematically proven non-statistical AI training model, ALIX, increased accuracy of finding patients who had cancer". Also, a "biproduct of the training identified and rank ordered the biomarkers from the most relevant to irrelevant. The glass box solution was explainable and repeatable". Such abilities will help in "early detection of cancers, increase precision medicine solutions for patients and treatment outcomes in the future" [78].

**Cost**

Three cost-saving AI solutions include virtual assist (chatbots), human assist (which routes complex customer questions to a human), and screen assist (which provides common answers to humans) [79].

These AI technologies can save millions of dollars for financially stressed businesses in today's challenging times by enabling them to address issues that affect customer service, costs and revenues [79].

AI has already increased productivity and efficiency in healthcare delivery, which has helped improve care outcomes, patient experiences and access to medical services [80].

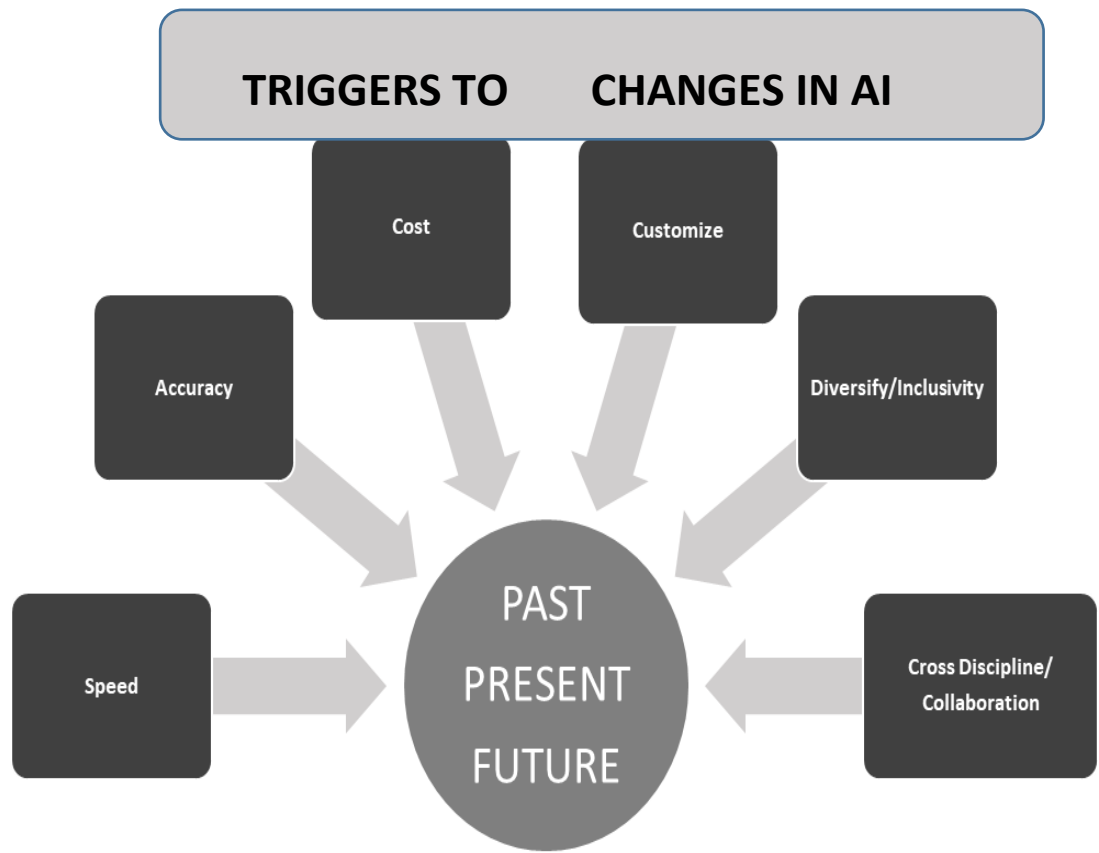

Figure 2: Triggers to Changes in Artificial Intelligence: Past, Present and Future

**Customization**

Customization is the name of the game – industries are using AI to humanize, personalize and customize products and services to their clients and expand their outreach and engagement [81]. Hyper personalization is the use of customer data to create and present customized contacts, information, or recommendations to customers. These customizations are created based on individual customer profiles. Profiles rely on data from browsing patterns, purchase histories, geographic location, demographic data, and behavioral data [82].

For example, Thread, a UK-based fashion retailer, offers customers AI-based product recommendations as a "personal stylist", from information collected from style quizzes and ongoing reactions to product recommendations and they do this with minimal additional effort or staffing [82].

Hilton Hotels currently uses a robot concierge named Connie in its lobbies to greet guests, answer questions and provide concierge-like answers to guests using natural language processing capabilities to interact with guests and develop meaningful profiles [82].

Although Under Armour is known for clothing, they reach customers through lifestyle activities of health and fitness, so they created the Record app, which collects user information on sleep, diet and physical activity. They then create personalized health goals and workout plans and after customers work out will provide feedback on the user's workout effectiveness to help maximize their future efforts [82].

**Diversity/Inclusivity**

While AI is quickly becoming a new tool in the CEO tool belt to drive revenues and profitability, it has also become clear that deploying AI requires careful management to prevent "unintentional but significant damage, not only to brand reputation but, more importantly, to workers, individuals, and society as a whole" [83, p1].

Recent research shows that AI bots and voice assistants promoted unfair gender stereotypes by featuring gendered names, voices, or appearances. In the United States, Siri, Alexa, Cortana, and Google Assistant—which collectively total an estimated 92.4% of U.S. market share for smartphone assistants—have traditionally featured female-sounding voices due to the designers' innate biases that female voices are more helpful, pleasant and accommodating than male ones [84]. In addition, racial and cultural biases also make it difficult for many people to interact easily with AI assistants around the world [85].

AI chatbots, recruitment software and risk assessment tools in the past caused harm by being racist, gender-biased or selecting the wrong people to put into jail [76]. People may not care how Facebook identifies who to tag in a given picture, but when AI systems are used to make diagnostic suggestions for skin cancer based on automatic picture analysis, understanding how such recommendations have been derived becomes a critical issue [63].

Experts say that AI is still "fragile, opaque, biased and not robust enough" to provide trustworthiness [87]. Leaders need to take the necessary steps to ensure that AI is being used in an ethical manner by consistent reliance on organizational values.

Three ways to accomplish this are: 1) Clarify how values translate into the selection of AI applications, 2) Provide guidance on definitions and metrics used to evaluate AI for bias and fairness, and 3) Prioritize organizational values [83]. Expanding the concept of AI to 'Responsible AI' is essential to ensure fairness, ethics, security/safety, privacy, transparency and accountability issues are considered [88].

"Business leaders may claim that diversity and inclusivity are core goals, but they then need to follow through in the people they hire and the products their companies develop" [19]. Ensuring minorities are well represented among both users and evaluators of AI will make AI more accessible and inclusive [88].

The covid-19 pandemic first discovered in 2019 has accelerated the need for the adoption of digital tools in education, particularly in the science, technology engineering and mathematics (STEM) arena. A majority of software developers are still males with only 25% women in the U.S. and minority racial groups are totally underrepresented in technology fields [89].

The goal is to create a stronger foundation for STEM literacy, inclusion, and diversity of STEM students and preparing the STEM workforce of the future. With the growing demand for advanced skill sets, educators can provide creative and more targeted learning rather than focusing on the repetitive tasks of creating problem sets. The net result is better learning outcomes for a wider group of students and requires collegial partnering, ongoing development, and thorough testing to implement [84].

**Collaboration and Cross Discipline**

"Creating differentiated experiences through personalization and immersive education will play a crucial role in the growth of remote learning," said Avasant's Research Leader, Pooja Chopra [53]. "Educational institutions should collaborate with EdTech companies and progressive service providers to accelerate digital transformation" [53].

Bibliometric studies that connect different disciplines are of growing interest in the analysis of the impact of AI synergies and their future within the research community. An example of this is a paper [91] which shows that the structure and model of the scientific production of researchers worldwide and the relationships between quality, references, and synergies among authors increases as collaboration across disciplines is applied. Multi-disciplinary research is vital to effective and natural human-robot connections as well [92].

Interdisciplinary research in artificial intelligence is a way to garner synergistic outcomes across industries from the AI field. To that end, researchers [93] recommend three strategies: 1) Collaborate on ways AI can impact other fields and look to new ideas from other fields to apply to AI; 2) Explain how decisions are made, be transparent about data biases, and use high level evaluators and regulators to evaluate processes; and 3) Scientific and educational experts should increase their AI educational levels.

Human-robot interaction challenges AI in many regards: dynamic, partially unknown environments not originally robot-friendly; a broad variety of situations with rich semantics to understand and interpret; human interactions requiring fine yet socially acceptable control strategies; natural and multi-modal communication requiring common-sense knowledge and divergent mental models. Collaboration of researchers and practitioners from across a variety of fields to integrate and share their data, knowledge, understandings and experiences is essential to meeting these challenges [58]. Cross-functional AI teams made up of diverse participants lead to greater innovations, more collaborations and better outcomes [94].

## CONCLUSIONS

In this paper outlining a 'Framework for Artificial Intelligence', the authors analyzed the triggers for AI development as well as the influencers to AI adoption. There is no doubt that current triggers such as speed, cost, accuracy, diversity/inclusion, competitiveness, personalization and the need for cross-disciplinary collaboration will continue into the foreseeable future. The present factors such as the coronavirus pandemic of 2019, climate change, customer needs, or resources may fluctuate or change in the future, but there will always be influencers that encourage wider AI adoption and those that discourage AI deployment in organizations. In this comprehensive look at the past, present and future applications in key industry sectors, a better and more comprehensive model for AI emerges.

AUTHORS

**Professor Richard A. Fulton** (M.S., Illinois State University) has taught full time computer science and information systems courses at Troy University – e campus for the past 18 years and previously at Illinois State University. His articles have been published in *The Journal of Technology Research, The Journal of Scientific Information on Political Theory, Developments in Business Simulations and Experiential Learning,* and the *International Journal of Innovation, Technology and Management.* 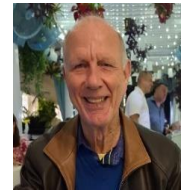

**Dr. Diane J. Fulton** (Ph. D., University of Tennessee-Knoxville) is Emeritus Professor of Management at Clayton State University, located in Morrow, Georgia. Her research interests include advanced technologies, innovations and online teaching tools. She has published several books, book chapters, and numerous articles in academic journals, including *California Management Review, Planning Review, Journal of Small Business Management, International Journal of Management Education and Entrepreneurship Theory and Practice.* 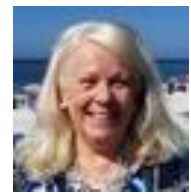

**Susan Kaplan** (BAS, MAS, University of Minnesota-Duluth) is Executive Vice President and Chief Management Officer of Modal Technology Corporation, a hightech firm located in Minneapolis, Minnesota that offers new and proven solutions for artificial intelligence and machine learning. Ms. Kaplan is a Founder and Director at RISC AI and was Founder and President for Quality Management Systems. She aided organizations in healthcare, government, manufacturing, and service sectors to improve profitability. She is the author of *The Grant Writing Process*. 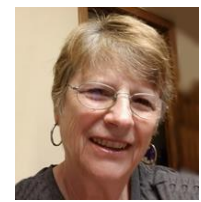